\documentclass[prb,twocolumn,superscriptaddress,showpacs]{revtex4-1}
\usepackage{graphicx}
\usepackage{hyperref}
\usepackage{multirow}
\usepackage{verbatim}
\usepackage{fontenc}
\usepackage{natbib}

\begin{document}

\title{Low-energy structures of zinc borohydride Zn(BH$_4$)$_2$}

\author{Tran Doan Huan}
\affiliation{Department of Physics, Universit\"{a}t Basel, Klingelbergstrasse 82, 4056 Basel, Switzerland}
\author{Maximilian Amsler}
\affiliation{Department of Physics, Universit\"{a}t Basel, Klingelbergstrasse 82, 4056 Basel, Switzerland}
\author{Vu Ngoc Tuoc}
\affiliation{Institute of Engineering Physics, Hanoi University of Science and Technology, 1 Dai Co Viet Road, Hanoi, Vietnam}
\author{Alexander Willand}
\affiliation{Department of Physics, Universit\"{a}t Basel, Klingelbergstrasse 82, 4056 Basel, Switzerland}
\author{Stefan Goedecker}
\email{stefan.goedecker@unibas.ch}
\affiliation{Department of Physics, Universit\"{a}t Basel, Klingelbergstrasse 82, 4056 Basel, Switzerland}

\date{\today}

\begin{abstract}
We present a systematic study of the low-energy structures of zinc borohydride, a crystalline material proposed for the hydrogen storage purpose. In addition to the previously proposed structures, many new low-energy structures of zinc borohydride are found by utilizing the minima-hopping method. We identify a new dynamically stable structure which belongs to the $I4_122$ space group as the lowest-energy phase of zinc borohydride at low temperatures. A low transition barrier between $I4_122$ and $P1$, the two lowest-lying phases of zinc borohydride is predicted, implying that a coexistence of low-energy phases of zinc borohydride is possible at ambient conditions.  An analysis based on the simulated X-ray diffraction pattern reveals that the $I4_122$ structure exhibits the same major features as the experimentally synthesized zinc borohydride samples.
\end{abstract}


\maketitle

\section{Introduction}

Hydrogen is an environment-friendly energy carrier which can provide a high energy density without producing greenhouse gases. This clean fuel is promising for many applications in, for example, the transportation sector. Various modern generations of fuel-cell and hybrid vehicles are being commercialized, using hydrogen as a fuel. Currently, hydrogen is stored in complex-structured high-pressure tanks of which the volume is a major difficulty for compact vehicles. A more efficient method for hydrogen storage is desired, and tremendous research efforts have been given to this goal \cite{mandal, review1, Pinkerton}.

Broad interest in metal borohydrides, a class of ionic crystal materials, is motivated by the possibility of using them for the hydrogen storage purpose \cite{review1,groch}. Among these materials, alkali metal borohydrides are generally thermodynamically too stable. For example, lithium borohydride LiBH$_4$ can reversibly store $8-10$\% hydrogen at temperatures of $315-400^\circ$C, which is too high for on-board applications \cite{review1}. Zinc borohydride Zn(BH$_4$)$_2$, one of the divalent metal borohydrides, is an alternative because of its more favorable thermodynamical properties. In particular, Zn(BH$_4$)$_2$ has a low decomposition temperature ($\simeq 85 ^\circ$C) \cite{jeon,groch,marks} and a relatively high gravimetric hydrogen density ($\simeq$ 8.5 wt \%) \cite{jeon,srini}. However, the reversibility of Zn(BH$_4$)$_2$ remains poorly understood. While Zn(BH$_4$)$_2$ was reported \cite{Srinivasan_WHEC2006} to be reversible, an attempt to reduce the decomposition temperature and to enhance the kinetic by doping Zn(BH$_4$)$_2$ with Ni nanoparticles suppressed the reversibility \cite{srini,srini2}. Considerable research interest was therefore devoted to this crystalline material \cite{marks, jeon, srini, srini2, Srinivasan_WHEC2006, naka06,chou,aidhy,chou2}.

Experimentally, Zn(BH$_4$)$_2$ can be synthesized in several ways \cite{marks,jeon,srini,srini2}, one of which is through the metathesis reaction
\begin{equation}\label{Eq:syn}
{\rm 2NaBH_4+ZnCl_2\to Zn(BH_4)_2+2NaCl}.
\end{equation}
Two X-ray powder diffraction (XRD) analysis for a Zn(BH$_4$)$_2$/NaCl mixture, the product of reaction (\ref{Eq:syn}), were reported in Refs. \onlinecite{jeon} and \onlinecite{srini}. However, the crystal structure of the Zn(BH$_4$)$_2$ products has not yet been conclusively determined because of the insufficient chemical purity of the samples \cite{jeon,srini}.

\begin{table*}[t]
\begin{center}
\caption{Summary of the known structures (top panel) and the new structures (bottom panel) of Zn(BH$_4$)$_2$. Enthalpies of formation at 0K ($\Delta H_{\rm 0K}$) and 100K ($\Delta H_{\rm 100K}$) as well as their components ($\Delta H_{\rm el}$, $\Delta H_{\rm ZP}$, and $\delta\Delta H_{\rm 100K}$) were obtained with the PBE functional. Total electronic energy $E^{\rm PBEsol}$ (without zero-point energy correction), obtained with the PBEsol functional, is given with respect to that of the $I4_122$ structure. The unit of the enthalpies and energies is kJ mol$^{-1}$ f.u.$^{-1}$. Space group (SG) numbers are given within the parentheses nest to the corresponding space group symbols.} \label{table_energy}
\begin{tabular}{l l l l r r r r r r}
    \hline
Initial material & Ref. & Initial SG & Final SG  &$\Delta H_{\rm el}$ & $\Delta H_{\rm ZP}$ & $\delta\Delta H_{\rm 100K}$ & $\Delta H_{\rm 0K}$
& $\Delta H_{\rm 100K}$ &$E^{\rm PBEsol}$\\
        \hline
Zn(BH$_4$)$_2$&[\onlinecite{aidhy}] & $F222$ (22)             & $I4_122$ (98)          & $-37.89$ & $3.11$ & $0.24$ & $-34.78$ & $-34.55$&0.00\\
Mg(BH$_4$)$_2$&[\onlinecite{zhou}]  &  $I4_122$ (98)          &$I4_122$ (98)           & $-37.89$ & $3.11$ & $0.24$ & $-34.78$ &$-34.55$&0.00\\
Zn(BH$_4$)$_2$&[\onlinecite{aidhy}] & $I\overline{4}m2$ (119) & $I\overline{4}m2$ (119)& $-36.32$ & $-$    & $-$     & $-$     &$-$     &1.64\\
Zn(BH$_4$)$_2$&[\onlinecite{naka06}] &$P\overline{1}$ (2)     & $P\overline{1}$ (2)    & $-26.45$ & $0.68$ & $0.06$ &  $-25.77$ &$-25.71$ &7.74\\
Zn(BH$_4$)$_2$&[\onlinecite{chou}] &$Pmc2_1$ (26)             & $Pmc2_1$ (26)          & $-19.06$ & $3.41$ & $0.71$ &  $-15.66$       &$-14.95$ &18.50\\
\hline
$-$&$-$& $-$ & $I4_122$ (98)           &  $-37.89$ & $3.11$  & $0.24$ & $-34.78$ & $-34.55$ & 0.00\\
$-$&$-$& $-$ & $I\overline{4}m2$ (119) & $-36.32$  & $-$     & $-$      & $-$    & $-$ &1.64\\
$-$&$-$& $-$ & $P1$ (1)                &  $-32.96$ & $2.23$  & $0.44$ & $-30.73$ & $-30.28$&5.92\\
$-$&$-$& $-$ & $C2$ (5)                & $-32.71$  & $1.80$  & $0.29$ & $-30.91$ & $-30.62$&6.11\\
$-$&$-$& $-$ & $Ibam$ (72)             & $-30.47$  & $1.90$  & $0.29$ & $-28.57$ & $-28.28$&8.30\\
$-$&$-$& $-$ & $C222_1$ (20)           & $-28.59$  & $-0.30$ & $-0.46$ & $-28.90$ &$-29.36$&14.92\\
$-$&$-$& $-$ & $Ama2$ (40)             &  $-27.44$ & $-1.49$ & $-0.86$ & $-28.92$ & $-29.78$&16.33\\
$-$&$-$& $-$ & $Pm$ (6)                &  $-27.41$ & $-1.47$ & $-0.83$ & $-28.88$ & $-29.72$&16.18\\
\hline
\end{tabular}
\end{center}
\end{table*}

Detailed knowledge of the crystal structure is however essential for further studies of a variety of material properties. Recently, ab-initio crystal structure prediction has become an increasingly attractive approach in material science, but still remains a challenging task (see Ref.~[\onlinecite{oganov_2010}] and references therein). Furthermore, density functional theory (DFT) \cite{dft1,dft2} calculations have been extensively used in the past to study hydrogen storage materials, and its successes and shortcomings have been discussed in detail by Herbst~\textit{et al.}~\cite{hector_enthalpy}. Several candidates for the low-temperature crystal structure of Zn(BH$_4$)$_2$ have been theoretically proposed in literature \cite{naka06,chou,aidhy}. All of the structures were predicted by the database searching method, starting from the existing crystal structures proposed for magnesium borohydride Mg(BH$_4$)$_2$, also a divalent borohydride. A triclinic $P\overline{1}$ structure was initially suggested from the monoclinic $P2/c$ structure predicted for Mg(BH$_4$)$_2$ in the same work by Nakamori {\it et al.} \cite{naka06} An orthorhombic $Pmc2_1$ structure was subsequently proposed and shown to be dynamically stable by Choudhury {\it et al.} \cite{chou}, starting from the $Pmc2_1$ structure initially intended for Mg(BH$_4$)$_2$~\cite{vaj}. Two other structures, also originally proposed for Mg(BH$_4$)$_2$, i.e., the tetragonal $I\overline{4}m2$ structure \cite{ozo08} and the orthorhombic $F222$ structure \cite{voss}, were then examined by Aidhy and Wolverton in Ref. \onlinecite{aidhy}, yielding two low-energy nearly-degenerated structures for Zn(BH$_4$)$_2$. While both of the structures are much lower in energy ($\simeq 30$ kJ mol$^{-1}$ f.u.$^{-1}$) than the $Pmc2_1$ structure, the $F222$ structure is energetically slightly favored over the $I\overline{4}m2$ structure by $\sim 1.0$ kJ mol$^{-1}$ f.u.$^{-1}$ (here f.u. is used for ``formula unit"). The $F222$ structure has therefore been used for the low-temperature phase of Zn(BH$_4$)$_2$ in a phase-stability analysis of some mixed-metal borohydride systems \cite{aidhy}.

In this paper, we revisit the low-temperature crystal structure of Zn(BH$_4$)$_2$ by first-principles calculations based on DFT. While all of the structures mentioned above are reexamined, we discover in addition a large number of low-energy structures of Zn(BH$_4$)$_2$ using the minima-hopping method \cite{goedecker04,amsler10}. The transition barrier between the two lowest-lying structures of Zn(BH$_4$)$_2$ is predicted by means of concerted nudged elastic band calculations. We then examine the dynamical stability of the obtained structures by phonon frequency calculations and analyze the simulated XRD patterns by comparing them with existing experimental results~\cite{jeon,srini}. Finally, we discuss some relevant features of the crystal structure prediction methods used for Zn(BH$_4$)$_2$.

\section{Computational methods}

First-principles calculations in this work were performed within the projector augmented wave formalism as implemented in the {\it Vienna Ab Initio Simulation Package} (\verb=VASP=) \cite{vasp1,vasp2,vasp3}. We used the generalized gradient approximation with the Perdew-Burke-Ernzerhof (PBE) functional \cite{per96} for the exchange and correlation energy. The semicore pseudopotential for zinc was used, of which the valence configuration is $3d^{10}4s^2$. For boron and hydrogen, the valence electron configurations are $2s^22p^1$ and $1s^1$, respectively. The convergence of the total energy calculations was ensured by a $9\times 9 \times 9$ Monkhorst-Pack $\bf k$-point mesh \cite{monkhorst} for sampling the Brillouin zone and a kinetic energy plane wave cutoff of 800 eV. Atomic and cell variables were simultaneously relaxed until all the residual force and stress components were smaller than 1 meV/\AA~and $10^{-3}$ kbar, respectively. The space groups corresponding to the relaxed structures were determined by \verb=FINDSYM= \cite{findsym}.

For searching the low-energy structures of Zn(BH$_4$)$_2$, we used the minima-hopping method\cite{goedecker04,amsler10}, an efficient structure prediction approach which was recently extended for crystalline systems. Different from other recent approaches which use, for example,  the electrostatic energy \cite{ozo08,maj08} or the number of metal-hydrogen bonds \cite{tekin10}, the minima-hopping method uses the energies evaluated at the DFT level as the objective function. The energy landscape is explored by short consecutive molecular dynamics trajectories followed by local geometry relaxations. The initial velocities for the molecular dynamics runs are chosen approximately along soft mode directions, allowing efficient escapes from local minima, and aiming towards the global minimum. This method was successfully applied in a wide range of material structure predictions \cite{hellmann_2007, roy_2009, bao_2009, willand_2010, de11, amsler_crystal_2011, Livas, amsler12}, and is not at all restricted to ionic hydride materials but can be applied to any system of given compositions with arbitrary boundary conditions. Some of the theoretically predicted structures, e.g., the neutral Si clusters with more than 12 atoms and four-fold coordinated defect in silicon, were recently confirmed by experiments \cite{haertelt,Markevich}.

The energetic ordering of the low-energy structures of Zn(BH$_4$)$_2$ was examined via the enthalpy of formation $\Delta H_T$, given by  \cite{herbst_enthalpy, hector_enthalpy}
\begin{equation}\label{eq:ef}
\Delta H_T = \Delta H_{\rm el} + \Delta H_{\rm ZP} + \delta\Delta H_T.
\end{equation}
Here, $\Delta H_{\rm el}$ and $\Delta H_{\rm ZP}$ are the electronic and zero-point energy differences between the products and the reactants, according to the reaction \ref{Eq:syn}, while $\delta \Delta H_T$ is the energy change from 0K to $T$. Within the harmonic approximation, $\Delta H_{\rm ZP}$ and $\delta \Delta H_T$ were straightforwardly computed from the results of the frozen-phonon calculations as described in Section \ref{sec:stability}. We used the $P4_2/mnc$ phase of NaBH$_4$ \cite{NaBH4}, the $Fm\overline{3}m$ phase of NaCl, and the $Pna2_1$ phase of $\delta-$ZnCl$_2$ \cite{ZnCl2} for the calculations of $\Delta H_T$ according to Eq. \ref{Eq:syn}.

\section{Low-temperature structure of zinc borohydride}
\subsection{Low-energy structures}

We re-examined the crystal structures previously proposed by Refs. \onlinecite{naka06}, \onlinecite{chou}, \onlinecite{aidhy} for Zn(BH$_4$)$_2$ or by Ref. \onlinecite{zhou} for Mg(BH$_4$)$_2$. Although the forces exerting on the ions of the $F222$ structure taken directly from Ref. \onlinecite{aidhy} are quite small ($\lesssim 0.05$ eV/\AA), one still can gain $\sim 1.7$ kJ mol$^{-1}$ f.u.$^{-1}$ by further relaxation. The relaxed structure was identified as an $I4_122$ structure, which can also be obtained by relaxing the $I4_122$ structure reported in Ref. \onlinecite{zhou} for Mg(BH$_4$)$_2$ after replacing the Mg atoms by Zn atoms. For the other structures, the corresponding space group remains unchanged after the relaxation.

\begin{table}[b]
\begin{center}
\caption{Atomic positions of the tetragonal $I4_122$ structure for Zn(BH$_4$)$_2$. Cell parameters are $a=b=6.986$\AA, $c=12.189$\AA, $\alpha=\beta=\gamma=90^\circ$. } \label{table_atpos}
\begin{tabular}{|l|l|r|r|r|}
\hline
Atom & Wyckoff site & $x$ & $y$ & $z$\\
\hline
Zn & 4a   &  0.0000 &  0.0000  & 0.0000\\
B  & 8f   & -0.0603 &  0.2500  & 0.1250\\
H  & 16g  & -0.0500 & -0.2476  & 0.2049\\
H  & 16g  &  0.1631 & -0.3890  & 0.1117\\
\hline
\end{tabular}
\end{center}
\end{table}

Several minima-hopping simulations were performed to search for additional low-energy structures of Zn(BH$_4$)$_2$. We found that the minima-hopping method is able to predict not only the existing $I4_122$ and $I\overline{4}m2$ structures, but also a large number of new low-energy structures for Zn(BH$_4$)$_2$. We consider in this work the structures which were discovered within the energy range of $\simeq 10$ kJ mol$^{-1}$ f.u.$^{-1}$ above the lowest-energy structure. Detailed information on these structures can be found in the supplemental material \cite{supplement}.

The enthalpies of formation of all the examined structures are shown in Table \ref{table_energy} together with their components, according to Eq. \ref{eq:ef}. At both 0K and 100K, the $I4_122$ structure (see Table \ref{table_atpos} for detailed information) is the most thermodynamically stable structure of Zn(BH$_4$)$_2$ . Table \ref{table_energy} also shows that the electronic energy $\Delta H_{\rm el}$ of the $I\overline{4}m2$ structure is slightly higher than that of the $I4_122$ structure by $\simeq 1.7$ kJ mol$^{-1}$ f.u.$^{-1}$. These two phases are energetically favored over the $P\overline{1}$ and $Pmc2_1$ phases by $\simeq 10$ and by $\simeq 20$ kJ mol$^{-1}$ f.u.$^{-1}$, respectively. These results, obtained with the PBE functional, are consistent with those obtained with the PW91 functional for the exchange-correlation energy \cite{aidhy}.

To confirm the energetic ordering, we performed additional calculations for the electronic energy by \verb=ABINIT= \cite{abinit1,abinit2}. Our calculations were carried out with the norm-conserving Hartwigsen-Goedecker-Hutter pseudopotential\cite{HGH},  a plane-wave cutoff energy of 60 Hatree ($\approx 1600$ eV), and the PBEsol functional, which is a modified PBE generalized gradient approximation that improves equilibrium properties of solids \cite{PBEsol}. The obtained results for the electronic energy $E^{\rm PBEsol}$, which are also shown in Table \ref{table_energy} with respect to that of the $I4_122$ structure, is consistent with the energetic ordering obtained with the PBE exchange-correlation functional.

\begin{figure}[b]
  \begin{center}
  \includegraphics[width=7 cm]{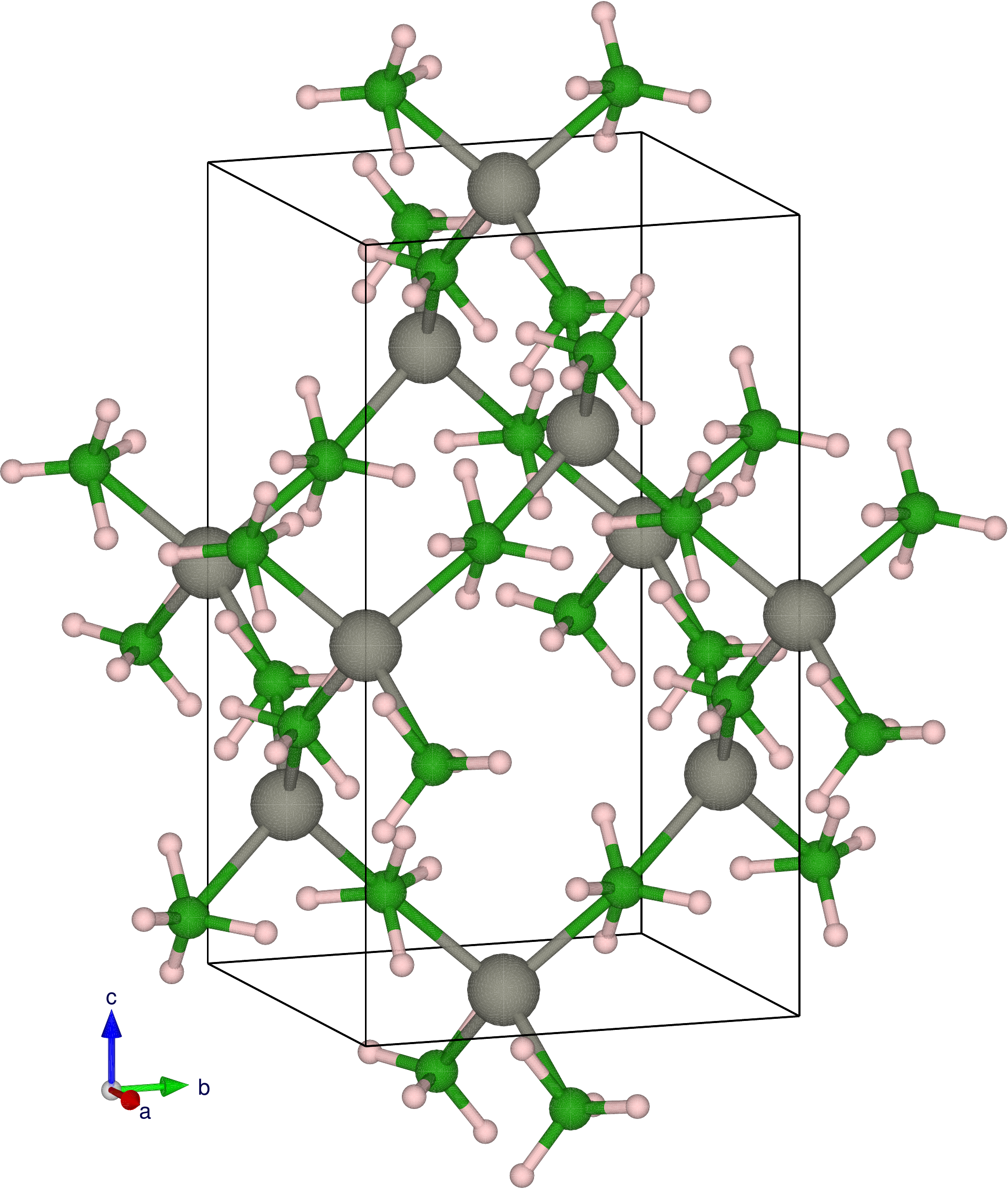}
  \caption{(Color online) The tetragonal $I4_122$ structure for Zn(BH$_4$)$_2$. Gray (large), green (medium), and pink (small) spheres represent zinc, boron, and hydrogen atoms, respectively.}\label{fig:struct}
  \end{center}
\end{figure}

The tetragonal $I4_122$ structure of Zn(BH$_4$)$_2$ is illustrated in Fig.~\ref{fig:struct} (prepared by \verb=VESTA= \cite{vesta}). Similar to the geometry of most of the other complex metal borohydrides, the complex [BH$_4$]$^-$ anions in the $I4_122$ structure for Zn(BH$_4$)$_2$ form isolated, slightly deformed tetrahedra with the B-H bond length of either 1.21\AA~ or 1.24\AA~ while the H-B-H angle is either $104^\circ$ or $119^\circ$. Each zinc atom is surrounded by four boron atoms with equal Zn-B bond length of 2.35 \AA~ and B-Zn-B angles of either $99^\circ$ or $134^\circ$. The geometries of the other structures shown in Table~\ref{table_energy} are somewhat similar, i.e., these ionic crystal structures are characterized by different arrangements of the Zn$^{2+}$ cations and the complex [BH$_4$]$^-$ anions in their more or less deformed tetrahedra.

\subsection{Structural stability}\label{sec:stability}

The dynamical stability of the zinc borohydride structures shown in Table \ref{table_energy} was determined by calculations of phonon frequencies using \verb=PHONOPY= \cite{phonopy}, a package based on the super-cell approach \cite{phonopy_sc}. For each relaxed structure, finite atomic displacements with an amplitude of 0.01 \AA~ were introduced to a $2 \times 2 \times 2$ super cell which contains 16 formula units of zinc borohydride (176 atoms). Calculations for the atomic forces within the super cells were then carried out by \verb=VASP=, allowing for the second-order force constants to be determined \cite{phonopy_sc}. The phonon frequencies of the structures were finally calculated from the dynamical matrices, given in terms of the force constants. The longitudinal optical/transverse optical (LO/TO) splitting was not taken into account since the effects of the LO/TO splitting were reported to be negligible for other hydrides \cite{amsler12,hector07,herbst10}.

\begin{figure}[t]
  \begin{center}
    \includegraphics[width=8.25cm]{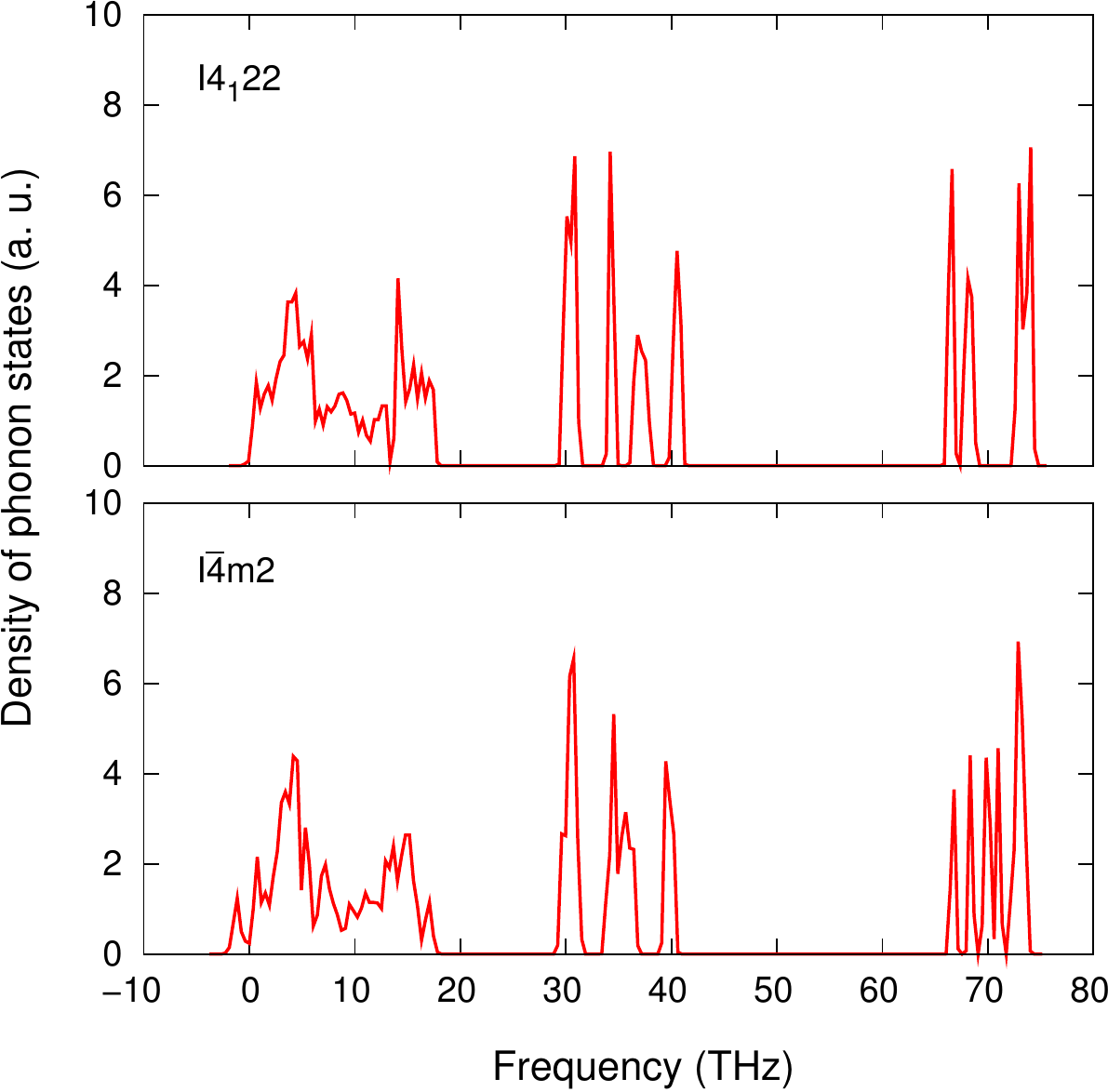}
  \caption{(Color online) Calculated density of phonon states of the $I4_122$ and $I\overline{4}m2$ structures for Zn(BH$_4$)$_2$. Imaginary phonon frequencies are represented on the figure by negative real values.}\label{fig:phonon_dos}
  \end{center}
\end{figure}

The obtained density of phonon states of the $I4_122$ structure, which is shown in Fig. \ref{fig:phonon_dos}, implies that this structure is dynamically stable. We also found that the $I\overline{4}m2$ structure is dynamically unstable because of two phonon modes, one of which presents at each of several high-symmetry points (see Fig. \ref{fig:phonon_dos} for the density of phonon states). In particular, at $Z$, $\Gamma$, $X$, and $N$, the corresponding imaginary mode has a frequency of $2.41i$ THz, $1.79i$ THz, $1.30i$ THz, and $1.19i$ THz, respectively. To explore these modes, we followed the corresponding atomic eigendisplacements at these points by the same procedure described in Refs. \onlinecite{herbst_phonon} and \onlinecite{sassi}. The aforementioned $I4_122$ structure was finally re-obtained by exploring the soft modes at either $Z$, $N$, or $X$ point. On the other hand, a dynamically stable structure which belongs to the $I\overline{4}$ space group (\# 82) was obtained by exploring the imaginary mode at $\Gamma$. However, the XRD pattern of the $I\overline{4}$ structure, which is slightly lower than the $I\overline{4}m2$ structure by 0.2 kJ mol$^{-1}$ f.u.$^{-1}$, is identical with that of the $I\overline{4}m2$ structure, clearly indicating that the Zn/B frames of these two structure are identical. Finally, the soft mode exploration of the $I\overline{4}m2$ structure ended up with the dynamically stable $I4_122$ structure.

The densities of phonon states of the other structures for Zn(BH$_4$)$_2$ are shown in the supplement material \cite{supplement}, indicating that they are dynamically stable. We note that the obtained density of phonon states for the $Pmc2_1$ structure, as shown in the supplement material\cite{supplement}, is consistent with the results reported in Ref. \onlinecite{chou}, which also indicates that the $Pmc2_1$ structure is dynamically stable.

\begin{figure*}[t]
  \begin{center}
  \includegraphics[width=0.86\textwidth]{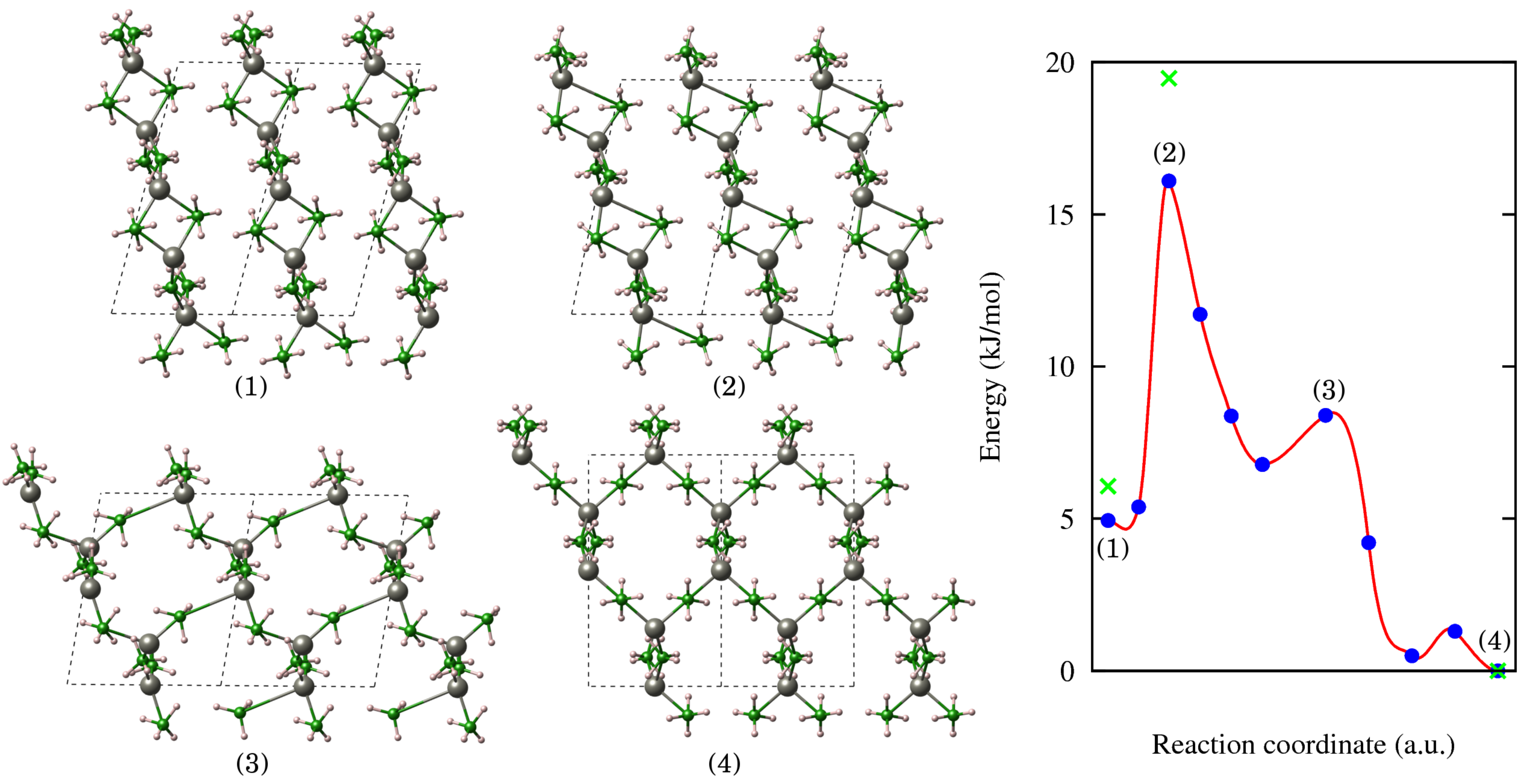}
  \caption{(Color online) Left panel: structural transition from the triclinic $P1$ (1) to the tetragonal $I4_122$ structure (4) for Zn(BH$_4$)$_2$ with intermediate steps. Gray (large), green (medium), and pink (small) spheres denote zinc, boron, and hydrogen atoms, respectively. Right panel: interpolated energies along the NEB pathway. Solid blue circles denote PBE results, whereas green crosses denote HSE06 energies. Structure at (2) corresponds to the saddle point.}\label{fig:neb}
  \end{center}
\end{figure*}

\subsection{Structural transformation}\label{sec:trans}
Polymorphism is commonly observed in molecular crystals and is driven by thermodynamically competing phases separated by low transition barriers. Upon a closer examination of Table~\ref{table_energy} we observe that a large number of structures exist in a small energy range of less than 10~kJ mol$^{-1}$ f.u.$^{-1}$ above the ground state. Therefore, coexistence or transformation of several different phases at finite temperature might be possible, especially if phase transitions require only small activation energies. A detailed and quantitative description of such a behaviour requires an accurate description of the free energy landscape, the kinetics of the transformation, the melting temperatures of each phase under consideration, etc. Here we shall limit ourselves to estimate an upper limit of the transition barrier between two selected dynamically stable phases, the triclinic $P1$ phase and the tetragonal $I4_122$ phase, at 0~K by employing the generalized solid-state nudged elastic band (G-SSNEB) method~\cite{sheppard_generalized_2012} as implemented in the VASP TST tools. An initial concerted pathway was estimated by carefully selecting an appropriate representation of the unit cells, followed by an approximative nudged elastic band (NEB) simulation. The saddle point was then refined by employing the climbing image NEB (CI-NEB) approach until the gradients were converged to less than 3~meV/\AA~at the first order transition state. The energies of the two end points and the saddle point structure were recomputed with higher accuracy using the HSE06 hybrid functional~\cite{heyd_hybrid_2003,paier_erratum:_2006,heyd_erratum:_2006} since PBE calculations are known to underestimates the barrier height, especially if the coordination number is reduced at the transition state~\cite{zupan_distributions_1997}.

The CI-NEB transition pathway is shown in Fig.~\ref{fig:neb} together with the structural evolution along the path. The planar structure in the $P1$ phase is distorted and recombined to form a hexagonal network in the $I4_122$ phase. The barrier height was found to be 11.16 kJ mol$^{-1}$ f.u.$^{-1}$ (reactant) and 16.10 kJ mol$^{-1}$ f.u.$^{-1}$ (product), while with HSE06 we found 13.40 kJ mol$^{-1}$ f.u.$^{-1}$ (reactant) and 19.47 kJ mol$^{-1}$ f.u.$^{-1}$ (product), respectively. These activation energies are roughly $2-3$ times larger than the energy differences between the two phases and are sufficiently low such that they could be easily overcome at ambient conditions. The phonon dispersion was calculated at the highest saddle point to confirm that a single imaginary phonon mode is present (see supplemental material for details). A second mode was identified  with partially imaginary frequencies close to the $\Gamma$ point. This indicates that the saddle point itself has very low curvatures with phonons contributing to a high rate constant according to transition state theory. Although we only investigated this particular phase transformation we can assume that similarly low barriers can be found for structural transitions between other phases since all such transformations do not require the breaking of strong covalent bonds but can be obtained by rearranging weakly bonded molecular subunits, as commonly observed in other molecular crystals.

\subsection{Structural identification}\label{sec:xrd}
While $I4_122$ was theoretically suggested to be the lowest-energy structure of Zn(BH$_4$)$_2$ at low temperatures, a comparison with available experimental data is particularly useful. Having some experimental XRD information for Zn(BH$_4$)$_2$ at hand~\cite{jeon,srini} we were able to perform such a comparative study. For this purpose, an XRD analysis was performed for all the examined structures using \verb=FULLPROF= package \cite{fullprof}. To be consistent with the reported experiments \cite{jeon,srini}, the Cu K$\alpha$ radiation (wavelength $\lambda = 1.54$\AA) was used for the XRD simulations. The simulated XRD patterns of these structures, given in the supplemental material \cite{supplement}, demonstrate that the $I4_122$ structure is different from the other structures, specifically the $F222$ structure.

\begin{figure}[b]
  \begin{center}
    \includegraphics[width=8.25 cm]{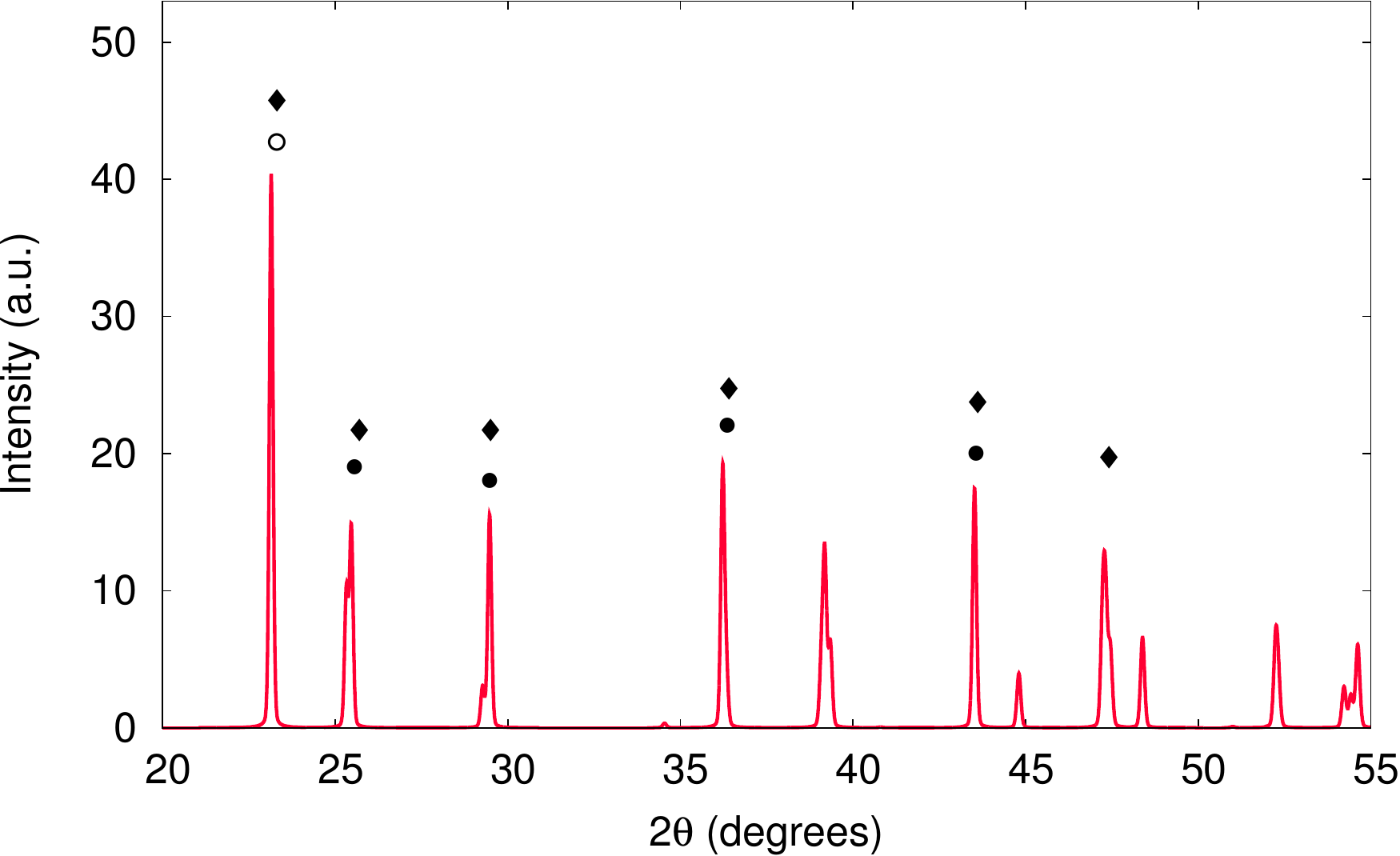}
  \caption{(Color online) Calculated XRD pattern of the $I4_122$ structure for Zn(BH$_4$)$_2$. Solid diamonds indicate the major peaks that were {\it not} claimed to correspond to NaCl in Ref. \onlinecite{jeon}. Solid (open) circles indicate the orientations of the major peaks that were (was not) claimed to correspond to the experimentally synthesized Zn(BH$_4$)$_2$ sample in Ref. \onlinecite{srini}. }\label{fig:xrd}
  \end{center}
\end{figure}

Our analysis on the simulated XRD patterns indicates that among the examined structures, the XRD pattern of the $I4_122$ matches favorably with the currently available XRD data \cite{jeon, srini}. In Fig. \ref{fig:xrd} we show the simulated XRD pattern of the $I4_122$ structure together with some information extracted from the experiments. As previously mentioned, two measured XRD patterns of the Zn(BH$_4$)$_2$/NaCl mixtures were reported in Refs. \onlinecite{jeon} and \onlinecite{srini}. In Ref. \onlinecite{jeon}, several dominant peaks corresponding to NaCl were identified while some of the remaining unlabeled peaks are assumed to correspond to Zn(BH$_4$)$_2$. More recently, several peaks from $25^\circ$ to $30^\circ$ and from $35^\circ$ to $45^\circ$ were explicitly suggested to correspond to Zn(BH$_4$)$_2$ by Ref. \onlinecite{srini}. In Fig. \ref{fig:xrd}, solid diamonds and solid circles indicate the positions of the peaks that were implied by Ref. \onlinecite{jeon} and \onlinecite{srini} to correspond to Zn(BH$_4$)$_2$. Although the information extracted from the experimentally observed XRD pattern is not sufficient for a conclusive structure determination, a qualitative discussion on the identification of the examined structures for Zn(BH$_4$)$_2$ is possible.

Fig. \ref{fig:xrd} indicates that the XRD pattern calculated for the $I4_122$ structure matches quite well with the XRD information extracted from Refs. \onlinecite{jeon} and \onlinecite{srini}. There is a major peak located at the orientation of $\simeq 23^\circ$ (indicated by an open circle), which was not identified to belong to Zn(BH$_4$)$_2$ in Ref. \onlinecite{srini} but can be seen in Ref. \onlinecite{jeon}. At this orientation of the experimental XRD pattern by Ref. \onlinecite{srini}, there is however an intense peak identified to correspond to the sample holder. A possible overlap of the characteristic peaks of Zn(BH$_4$)$_2$ and the sample holder may be the reason explaining why this major peak of the Zn(BH$_4$)$_2$ crystal was not explicitly identified \cite{srini}.

Fig. \ref{fig:xrd}  demonstrates that the $I4_122$ structure shares major structural features with that of the experimentally synthesized Zn(BH$_4$)$_2$ samples \cite{jeon,srini}. Obviously, the experimentally measured XRD patterns for the Zn(BH$_4$)$_2$/NaCl mixtures can not provide sufficiently accurate information for determining the low-temperature crystal structure of Zn(BH$_4$)$_2$. Therefore, XRD data for purified crystalline Zn(BH$_4$)$_2$ would be desirable for a proper determination of the structure.

\begin{table}[t]
\begin{center}
\caption{Total electronic energy $E^{\rm PBE}$ and dynamical stability of the examined structures for Zn(BH$_4$)$_2$ and Mg(BH$_4$)$_2$. The energy is given in unit of kJ mol$^{-1}$ f.u.$^{-1}$ with respect to that of the $I4_122$ structure.} \label{table_energy_comp}
\begin{tabular}{|l|r|r|r|r|}
\hline
\multirow{2}{*}{Structure} & \multicolumn{2}{|c|}{Zn(BH$_4$)$_2$} & \multicolumn{2}{|c|}{Mg(BH$_4$)$_2$}\\
\cline{2-5}
& $E^{\rm PBE}$ & Stability & $E^{\rm PBE}$ &Stability\\
\hline
$I4_122$ & 0.00 & stable   & 0.00& stable \\
$I\overline{4}m2$ & 1.68 & unstable & 0.10 & unstable \\
$C222_1$ & 9.30& stable & 57.33 & unstable\\
$Ama2$ & 10.45 & stable & 33.97 & unstable\\
$P\overline{1}$ & 11.57 & stable & 31.27 & stable\\
$Pmc2_1$ & 18.84 & stable & 21.69 &stable \\
\hline
\end{tabular}
\end{center}
\end{table}

\section{Crystal structure prediction methods for zinc borohydride}
Prior to our investigations all the proposals for ionic crystal structures of Zn(BH$_4$)$_2$ were based on those of Mg(BH$_4$)$_2$ \cite{naka06,chou,aidhy}. It was also found \cite{aidhy} that the energy ordering of the $F222$, $I\overline{4}m2$, and $Pmc2_1$ structures for Zn(BH$_4$)$_2$ is similar to the corresponding energy ordering of Mg(BH$_4$)$_2$, which phenomenologically shows a structural correspondence between Zn(BH$_4$)$_2$ and Mg(BH$_4$)$_2$. This observation is also supported by the so-called Goldschmidt's rules of substitution \cite{goldschmidt}, according to which one can substitute Mg atoms in an ionic crystal structure of Mg(BH$_4$)$_2$ by Zn atoms without disrupting the structural stability due to the similar charges and ionic radii of the Zn$^{2+}$ and the Mg$^{2+}$ cations.

To further explore this structural correspondence, we determined the energy ordering and the dynamical stability of an extended list of structures, including three more structures obtained in this work, i.e., the $P\overline{1}$, $C222_1$, and $Ama2$ structures. For each of the structures, which were already obtained for Zn(BH$_4$)$_2$, Zn atoms were substituted by Mg atoms, then the cell and the atomic variables were fully relaxed. As reported by Ref. \onlinecite{zhou} and similar to the case of Zn(BH$_4$)$_2$, the $I4_122$ structure for Mg(BH$_4$)$_2$ was also obtained by relaxing the $F222$ structure. Phonon frequency calculations were then carried out to determine the dynamical stability of the structures.  A summary of this investigation is given in Table \ref{table_energy_comp} and Fig. \ref{fig:diag} while the densities of phonon states of these structures for Mg(BH$_4$)$_2$ are given in the supplemental material\cite{supplement}.

Table \ref{table_energy_comp} and Fig. \ref{fig:diag} show that, in agreement with Ref. \onlinecite{aidhy}, the energy ordering of the $I4_122$, $I\overline{4}m2$, and $Pmc2_1$ structures for Zn(BH$_4$)$_2$ is similar to that for Mg(BH$_4$)$_2$. However, the energy ordering of the extended list of structures for Zn(BH$_4$)$_2$ is different from that for Mg(BH$_4$)$_2$. Regarding the dynamical stability, four structures ($I4_122$, $I\overline{4}m2$, $P\overline{1}$ and $Pmc2_1$) exhibit the same behavior for Zn(BH$_4$)$_2$ and Mg(BH$_4$)$_2$ while the other two structures ($Ama2$ and $C222_1$) do not. Obviously, the phenomenological structural correspondence between Zn(BH$_4$)$_2$ and Mg(BH$_4$)$_2$ is weak, especially for the low-energy phases which are closed in energy.

\begin{figure}[t]
  \begin{center}
    \includegraphics[width=7.5 cm]{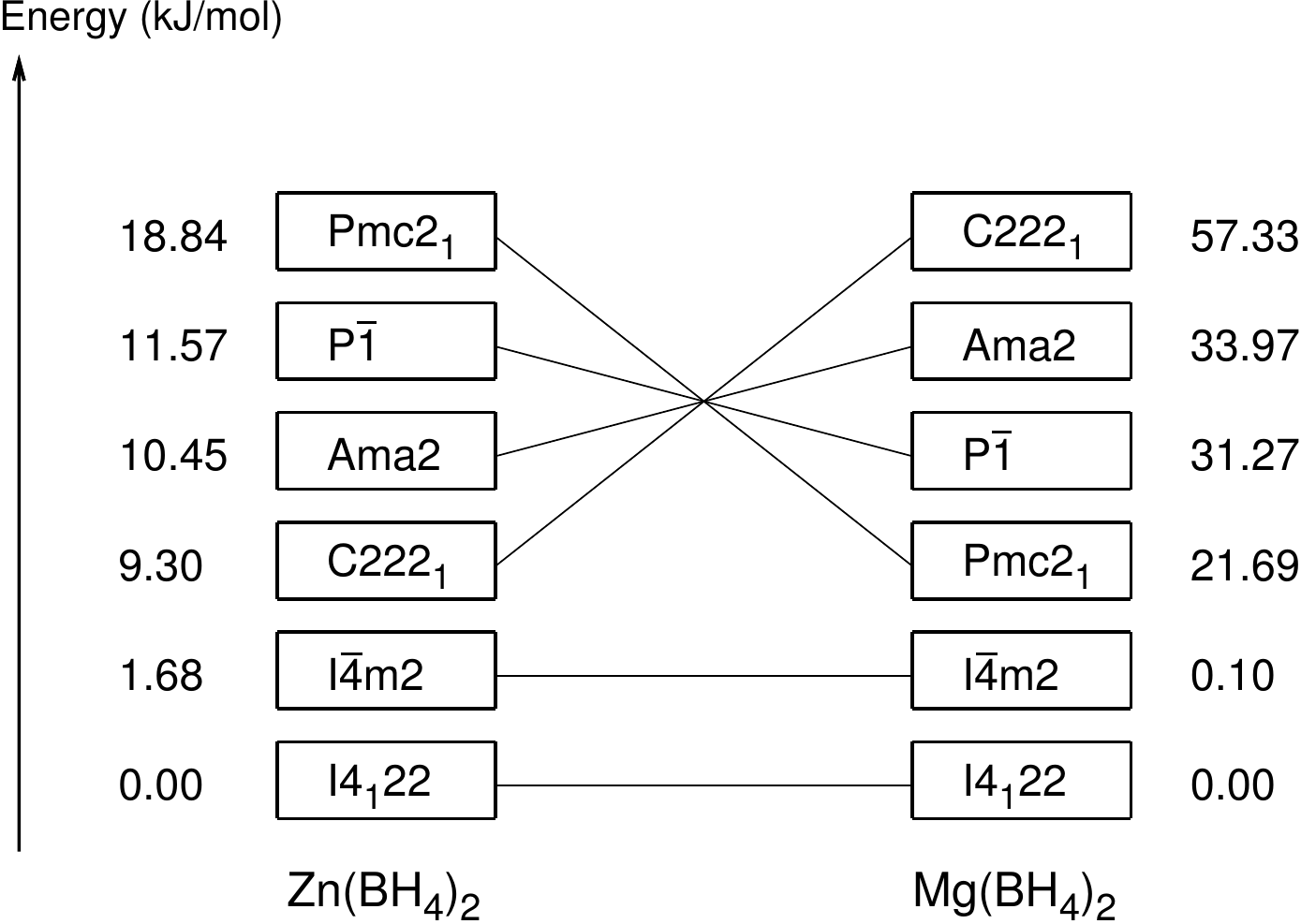}
  \caption{Energy ordering of the extended list of structures examined on Table \ref{table_energy_comp} for Zn(BH$_4$)$_2$ and Mg(BH$_4$)$_2$. Each structure is represented by a box with the corresponding space group. Numbers are given next to the boxes, indicating the energy difference (in kJ mol$^{-1}$ f.u.$^{-1}$) between the corresponding structures and the most stable structure. Lines connect the same structure for the two materials, indicating the re-arrangement of the examined structures when Zn atoms are substituted by Mg atoms. }\label{fig:diag}
  \end{center}
\end{figure}

It can be observed from Table \ref{table_energy} and also in Refs. \onlinecite{naka06}, \onlinecite{chou}, and \onlinecite{aidhy} that the symmetry is not likely to break if a local geometry relaxation is started from a structure obtained by substituting Mg by Zn in a Mg(BH$_4$)$_2$ phase. Consequently, in principle, it is hard to explore the new symmetries which have not been reported in the literature for Zn(BH$_4$)$_2$ and related materials, e.g., Mg(BH$_4$)$_2$. This issue is however solved by an unconstrained searching method, e.g., the minima-hoping method. As illustrated in Table \ref{table_energy}, one can easily explore the new symmetries which have not been reported.

It is worth noting that, while several previous studies \cite{ozo08,voss} reported that the $Pmc2_1$ structure for Mg(BH$_4$)$_2$ is dynamically unstable, the $Pmc2_1$ structure examined in this work for Mg(BH$_4$)$_2$ was determined to be dynamically stable. To clarify this discrepancy, we have performed additional calculations for the $Pmc2_1$ structure that was explicitly reported in Ref. \onlinecite{voss}. We found that, in agreement with the previous studies \cite{ozo08,voss}, this structure is indeed dynamically unstable. On the other hand, the $Pmc2_1$ structure examined in this work, taken from Ref. \onlinecite{chou} for Zn(BH$_4$)$_2$, is slightly deformed from that reported in Ref. \onlinecite{voss}, and using it for Mg(BH$_4$)$_2$ yields a dynamically stable structure. For more information, the densities of phonon states of these slightly deformed $Pmc2_1$ structures for Mg(BH$_4$)$_2$ are shown in the supplemental material \cite{supplement}.

\section{Conclusions}
In conclusion, we have carried out a systematic study of the low-energy structural phases of Zn(BH$_4$)$_2$. By using the minima-hopping method, we have discovered many new low-energy structures of Zn(BH$_4$)$_2$. The most stable structure is identified to belong to the $I4_122$ space group. Phonon calculations demonstrate that the $I4_122$ structure is dynamically stable while the $I\overline{4}m2$ structure, which is slightly higher in energy than the $I4_122$ structure, is dynamically unstable. By following the atomic eigendisplacements corresponding to the unstable phonon modes, the aforementioned dynamically stable $I4_122$ structure is finally re-obtained. An XRD analysis implies that the $I4_122$ structure shares some major structural features with the structure of the Zn(BH$_4$)$_2$ samples that were experimentally synthesized.

Furthermore, because many thermodynamically competing low-energy structures were found at low-energy during our structural search, and a concerted phase transition between the two lowest-lying structures, i.e., the $P1$ and the $I4_122$ structures, was investigated. We found that barrier connecting these two structures is low, implying that  Zn(BH$_4$)$_2$ might exhibit a polymorphic behavior at ambient conditions.

The structural similarity between Zn(BH$_4$)$_2$ and Mg(BH$_4$)$_2$, mentioned in Ref. \onlinecite{aidhy}, plays an important role when proposing low-energy structures of Zn(BH$_4$)$_2$. We have discussed the similarity and found that, although there is a certain structural correspondence between Zn(BH$_4$)$_2$ and Mg(BH$_4$)$_2$, it is weak and strongly limits the exploration of new symmetries of Zn(BH$_4$)$_2$ at low energies. We show that the minima-hopping method, on the other hand, allows for an efficient and fully unconstrained structural search of Zn(BH$_4$)$_2$.

\begin{acknowledgments}
The authors thank D. S. Aidhy, Nguyen-Manh Duc, S. Alireza Ghasemi, Jos\'e A. Flores-Livas, Dam Hieu Chi, and A. Tekin for useful discussions and correspondence. They also thank the referees for useful comments and suggestions. TDH, MA, AW, and SG gratefully acknowledge the financial support provided by the Swiss National Science Foundation. Work by VNT is supported by the Vietnamese NAFOSTED program No. 103.02-2011.20. TDH and MA acknowledge the computational resources provided by the Swiss National Supercomputing Center (CSCS) in Manno and Lugano, Switzerland.
\end{acknowledgments}

\end{document}


\title{Supplemental material: Low-energy structures of zinc borohydride Zn(BH$_4$)$_2$}

\author{Tran Doan Huan}
\affiliation{Department of Physics, Universit\"{a}t Basel, Klingelbergstrasse 82, 4056 Basel, Switzerland}
\author{Maximilian Amsler}
\affiliation{Department of Physics, Universit\"{a}t Basel, Klingelbergstrasse 82, 4056 Basel, Switzerland}
\author{Vu Ngoc Tuoc}
\affiliation{Institute of Engineering Physics, Hanoi University of Science and Technology, 1 Dai Co Viet Road, Hanoi, Vietnam}
\author{Alexander Willand}
\affiliation{Department of Physics, Universit\"{a}t Basel, Klingelbergstrasse 82, 4056 Basel, Switzerland}
\author{Stefan Goedecker}
\affiliation{Department of Physics, Universit\"{a}t Basel, Klingelbergstrasse 82, 4056 Basel, Switzerland}

\date{\today}

\maketitle

\begin{longtable}{|l|l|}
\caption{Calculated parameters for the structures discovered by the minima-hopping method 
for zinc borohydride. For each structure, cell parameters are given while 
for each atom, the Wyckoff site and the coordinates ($x$, $y$, and $z$) are given.} \\
\endfirsthead

\multicolumn{2}{c}%
{{\bfseries \tablename\ \thetable{} -- continued from previous page}} \\
\hline

\endhead

\hline 
\multicolumn{2}{|r|}{{to be continued .. }} \\ 
\hline
\endfoot

\hline \hline
\endlastfoot

\hline 
\hline
\hline
$P1$ & $a=7.37$\AA, $b=6.96$\AA, $c=6.42$\AA~\\
  (1) & $\alpha=71.47^\circ$, $\beta=109.31^\circ$, $\gamma=118.14^\circ$\\
\hline
Zn & (1a) (-0.2109, -0.3116, 0.3580)\\
Zn & (1a) (0.2494, 0.4200, 0.3545)\\
B  & (1a) (0.0104, 0.0688, 0.2853)\\
B  & (1a) (0.4698, -0.3169, 0.0886)\\
B  & (1a) (0.0262, -0.4655, 0.4222)\\
B  & (1a) (-0.4210, 0.4392, -0.3726)\\
H  & (1a) (0.1871, 0.1170, 0.4143)\\
H  & (1a) (0.3214, 0.4931, 0.0885)\\
H  & (1a) (0.0100, -0.0349, 0.1638)\\
H  & (1a) (0.4313, -0.1989, -0.0870)\\
H  & (1a) (0.4251, 0.4745, -0.3780)\\
H  & (1a) (-0.4436, 0.3551, 0.4745)\\
H  & (1a) (0.1569, -0.3587, 0.3003)\\
H  & (1a) (0.0229, 0.3583, -0.4576)\\
H  & (1a) (-0.2764, -0.3701, -0.3687)\\
H  & (1a) (-0.3873, 0.3204, -0.1956)\\
H  & (1a) (0.0707, -0.3406, -0.4539)\\
H  & (1a) (-0.1452, 0.4872, 0.2852)\\
H  & (1a) (-0.1116, -0.0283, 0.4167)\\
H  & (1a) (-0.4959, -0.2305, 0.2439)\\
H  & (1a) (-0.0408, 0.2180, 0.1643)\\
H  & (1a) (-0.3831, -0.3565, 0.0822)\\
\hline
\hline
\hline
$C2$ & $a=6.98$\AA, $b=12.98$\AA, $c=6.29$\AA~\\
  (5) & $\alpha=90^\circ$, $\beta=71.67^\circ$, $\gamma=90^\circ$\\
\hline
Zn &  (2a) (0.0000, -0.2392, 0.0000) \\
Zn &  (2a) (0.0000, 0.0300, 0.0000)\\
B  &  (4c) (-0.3533, 0.3978, -0.2684)\\
B  &  (4c) (-0.2703, 0.1453, 0.0684)\\
H  &  (4c) (-0.4676, 0.3226, -0.2657)\\
H  &  (4c) (-0.2312, 0.2218, -0.0570)\\
H  &  (4c) (0.2193, 0.3964, 0.4442)\\
H  &  (4c) (-0.4433, 0.1316, 0.1885)\\
H  &  (4c) (-0.2804, 0.4003, -0.1153)\\
H  &  (4c) (-0.2342, 0.0719, -0.0648)\\
H  &  (4c) (0.4686, 0.4711, 0.2733)\\
H  & (4c) (-0.1660, 0.1565, 0.1903)\\
\hline
\hline
\hline
$Pm$ & $a=6.74$\AA, $b=8.00$\AA, $c=5.67$\AA~\\
   (6)      & $\alpha=90^\circ$, $\beta=67.21^\circ$, $\gamma=90^\circ$\\
\hline
Zn & (1b) (0.3547, 0.5000, -0.2154)\\
Zn & (1a) (0.1220, 0.0000, -0.1081)\\
B  & (1b) (-0.3953, 0.5000, 0.4045)\\
B  & (2c) (0.2376, -0.2502, -0.0458)\\
B  & (1a) (-0.1276, 0.0000, -0.2582)\\
H  & (1b) (-0.3596, 0.5000, -0.3896)\\
H  & (2c) (0.1042, -0.1862, 0.1477) \\
H  & (2c) (-0.1996, 0.1301, -0.2977) \\
H  & (2c) (0.1301, 0.3438, -0.1254) \\
H  & (2c) (0.3698, -0.3149, 0.0262) \\
H  & (1b) (0.4062, 0.5000, 0.4433) \\
H  & (2c) (-0.3229, -0.3699, 0.2990) \\
H  & (1a) (0.0711, 0.0000, -0.4021) \\
H  & (2c) (0.3467, -0.1560, -0.2230) \\
H  & (1a) (-0.1639, 0.0000, -0.0193) \\
\hline
\hline
\hline
$C222_1$ & $a=11.85$\AA, $b=6.65$\AA, $c=7.15$\AA~\\
   (20)      & $\alpha=\beta=\gamma=90^\circ$\\
\hline
Zn    &(4a) (0.3879, 0.0000, 0.0000)\\
B     &(4a) (0.2061, 0.0000, 0.0000)\\
B     &(4b) (0.0000, 0.4570, 0.2500)\\
H     &(8c) (0.2654, 0.1605, -0.0043)\\
H     &(8c) (-0.4890, 0.1579, -0.1160)\\
H     &(8c) (0.3453, 0.4960, -0.3546)\\
H     &(8c) (0.0833, 0.4313, -0.2438)\\
\hline
\hline
\hline
$Ama2$ & $a=8.00$\AA, $b=12.44$\AA, $c=5.67$\AA~\\
(40)   & $\alpha=\beta=\gamma=90^\circ$\\
\hline
Zn    &(4b) (0.2500, 0.0582, -0.1618)\\
B     &(4b) (0.2500, 0.1831, -0.4266)\\
B     &(4a) (0.0000, 0.0000, -0.0462)\\
H     &(4b) (0.2500, 0.2010, -0.2044)\\
H     &(8c) (-0.0643, -0.0664, 0.0861)\\
H     &(8c) (-0.3802, 0.2809, 0.0010)\\
H     &(8c) (0.4061, 0.4459, 0.3258)\\
H     &(4b) (0.2500, 0.0838, -0.4791)\\
\hline
\hline
\hline
$Ibam$ & $a=7.68$\AA, $b=9.25$\AA, $c=6.84$\AA~\\
  (72) & $\alpha=\beta=\gamma=90^\circ$\\
\hline
Zn & (4b) (0.5000, 0.0000, 0.2500)\\
B  & (8j) (0.3069, 0.0847, 0.0000)\\
H  & (8j) (0.1559, 0.1189, 0.0000)\\
H  & (8j) (-0.1836, -0.4524, 0.0000)\\
H  & (16k) (0.1200, -0.3525, 0.1406)\\
\hline
\end{longtable}

\begin{figure*}[ht]
  \begin{center}
    \includegraphics[width=16cm]{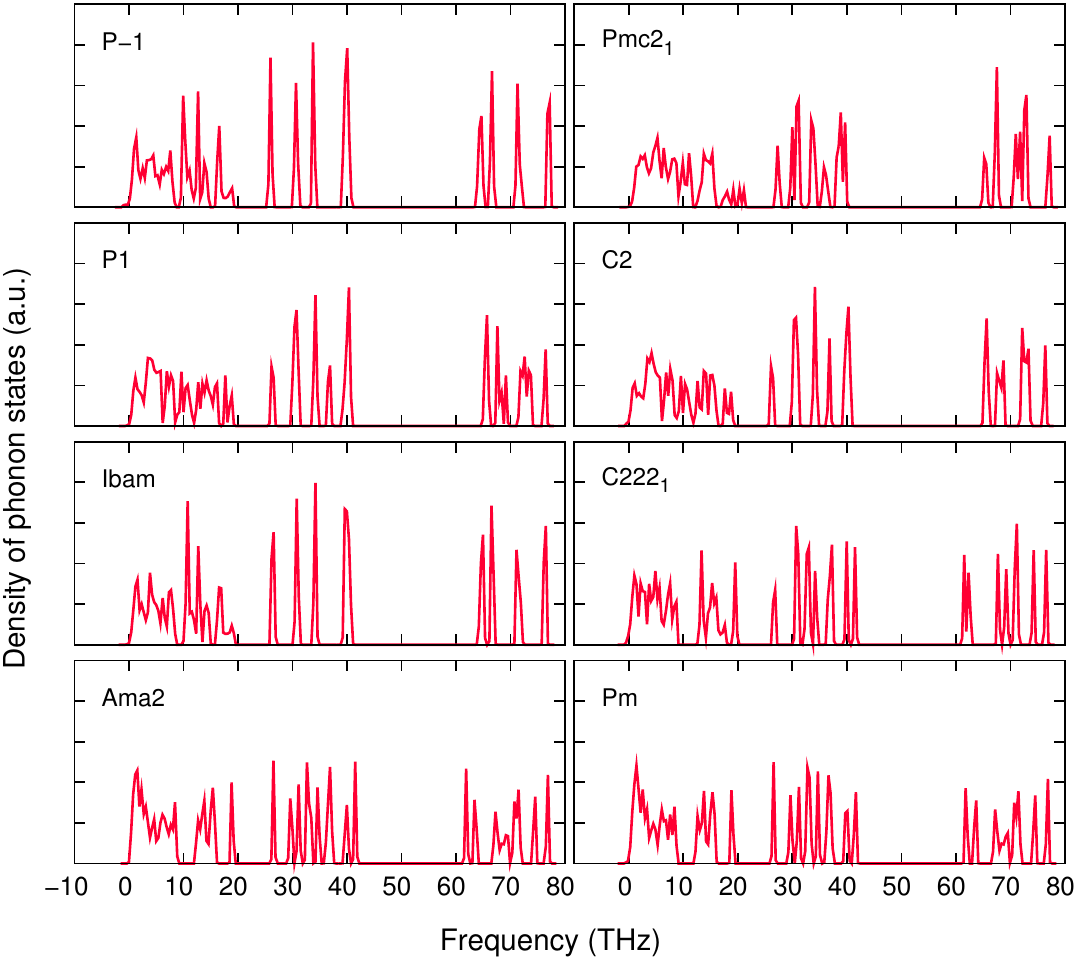}
  \caption{(Color online) Calculated density of phonon states of the $P\overline{1}$, $Pmc2_1$, $P1$, $C2$, $Ibam$, $C222_1$, $Ama2$, and $Pm$ structures for Zn(BH$_4$)$_2$.}\label{fig:phdosZn}
  \end{center}
\end{figure*}

\begin{longtable}{|l|l|}
\caption{Crystallographic information of the $P1$ saddle-point structure on the CI-NEB transition pathway of Zn(BH$_4$)$_2$. Cell parameters are given while 
for each atom, the Wyckoff site and the coordinates ($x$, $y$, and $z$) are given.} \\
\endfirsthead

\multicolumn{2}{c}%
{{\bfseries \tablename\ \thetable{} -- continued from previous page}} \\
\hline

\endhead

\hline 
\multicolumn{2}{|r|}{{to be continued .. }} \\ 
\hline
\endfoot

\hline \hline
\endlastfoot

\hline
\hline
$P1$ & $a=7.08$\AA, $b=6.98$\AA, $c=6.64$\AA~\\
  (1) & $\alpha=78.12^\circ$, $\beta=111.71^\circ$, $\gamma=113.67^\circ$\\
\hline
Zn & (1a) (0.0724, -0.0936, -0.2237)\\
Zn & (1a) (-0.3993, 0.2378, -0.3221)\\
B  & (1a) (-0.1833, -0.4233, -0.3321)\\
B  & (1a) (0.2446,  0.1748, -0.4899)\\
B  & (1a) (-0.2020,  0.0344, -0.2072)\\
B  & (1a) (0.3805, -0.0528,  0.0573)\\
H  & (1a) (0.0036, -0.3896, -0.2052)\\
H  & (1a) (0.1189, -0.0119, -0.4889)\\
H  & (1a) (-0.2015, 0.4616, -0.4623)\\
H  & (1a) (0.1535, 0.2913, 0.3999)\\
H  & (1a) (0.2130, -0.0229, 0.0522)\\
H  & (1a) (0.3561, -0.1333, -0.1055)\\
H  & (1a) (-0.0499, 0.1133, -0.2841)\\
H  & (1a) (-0.2205, -0.1271, -0.0944)\\
H  & (1a) (-0.4794, 0.1212, 0.0677)\\
H  & (1a) (0.4087, -0.1771, 0.2109)\\
H  & (1a) (-0.1694, 0.1680, -0.0887)\\
H  & (1a) (-0.3689, -0.0050, -0.3690)\\
H  & (1a) (-0.3007, -0.4975, -0.2140)\\
H  & (1a) (0.3376, 0.2399, -0.3043)\\
H  & (1a) (-0.2342, -0.2787, -0.4424)\\
H  & (1a) (0.3669, 0.1636, 0.4207)\\
\hline
\end{longtable}

\begin{figure*}[ht]
  \begin{center}
    \includegraphics[width=9cm]{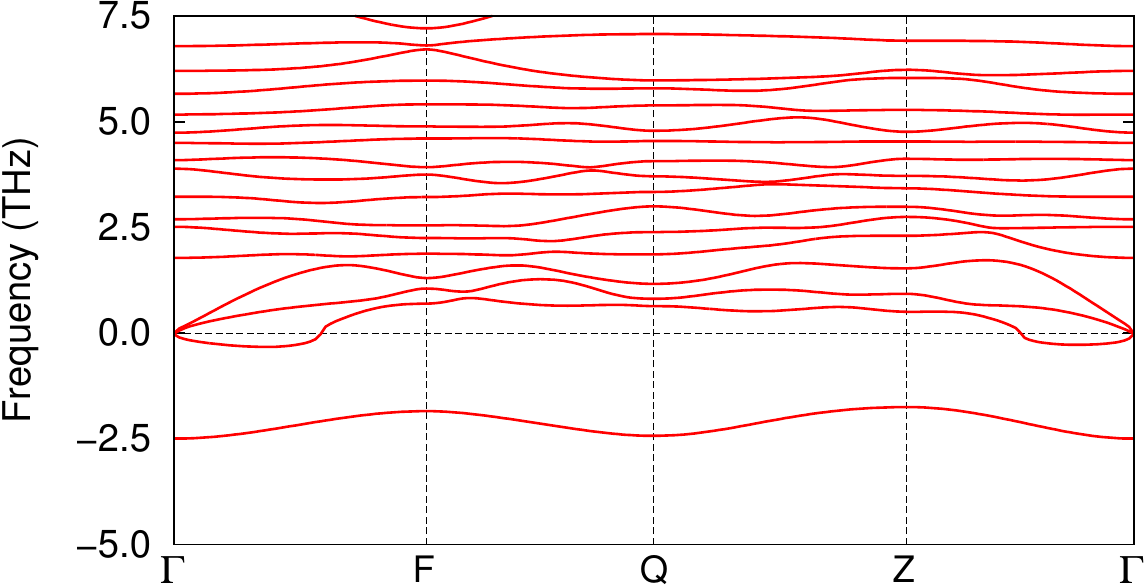}
  \caption{(Color online) Calculated density of phonon states of the highest ($P1$) saddle point structure on the CI-NEB transition pathway of Zn(BH$_4$)$_2$. See Fig. 3 of the text for more information.}\label{fig:phdosSaddle}
  \end{center}
\end{figure*}

\begin{figure*}[ht]
  \begin{center}
    \includegraphics[width=16cm]{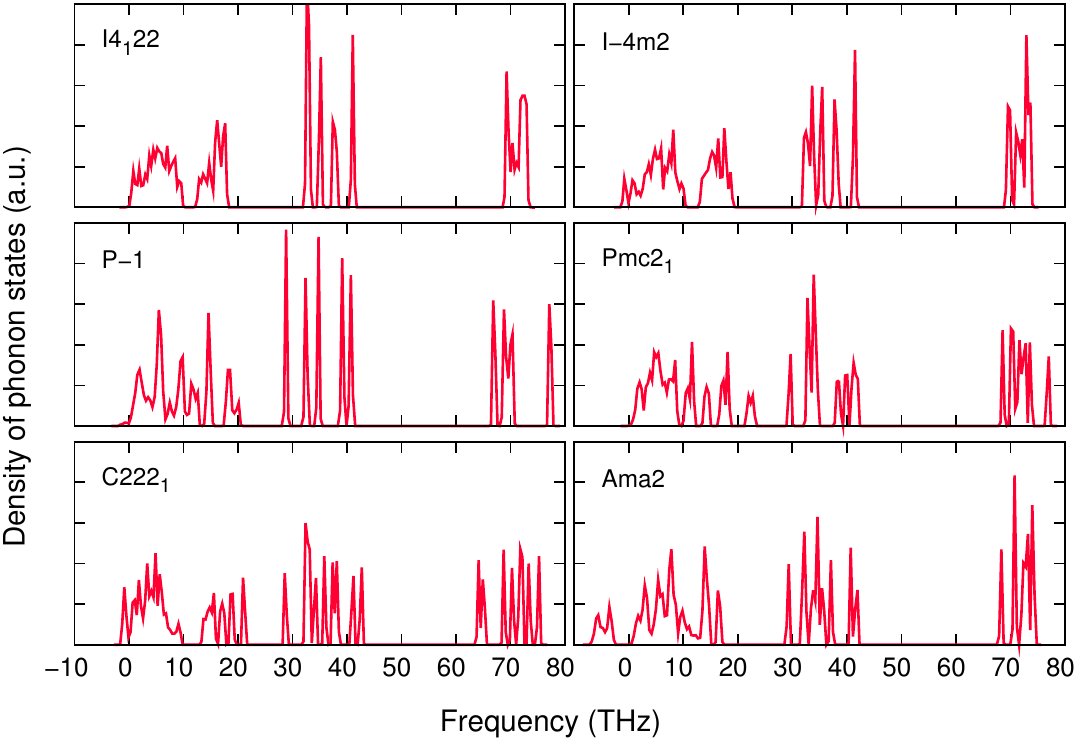}
  \caption{(Color online) Calculated density of phonon states of the $I4_122$, $I\overline{4}m2$, $P\overline{1}$, $Pmc2_1$, $C222_1$, and $Ama2$ structures for Mg(BH$_4$)$_2$.}\label{fig:phdosMg}
  \end{center}
\end{figure*}

\begin{figure}[ht]
  \begin{center}
    \includegraphics[width=11cm]{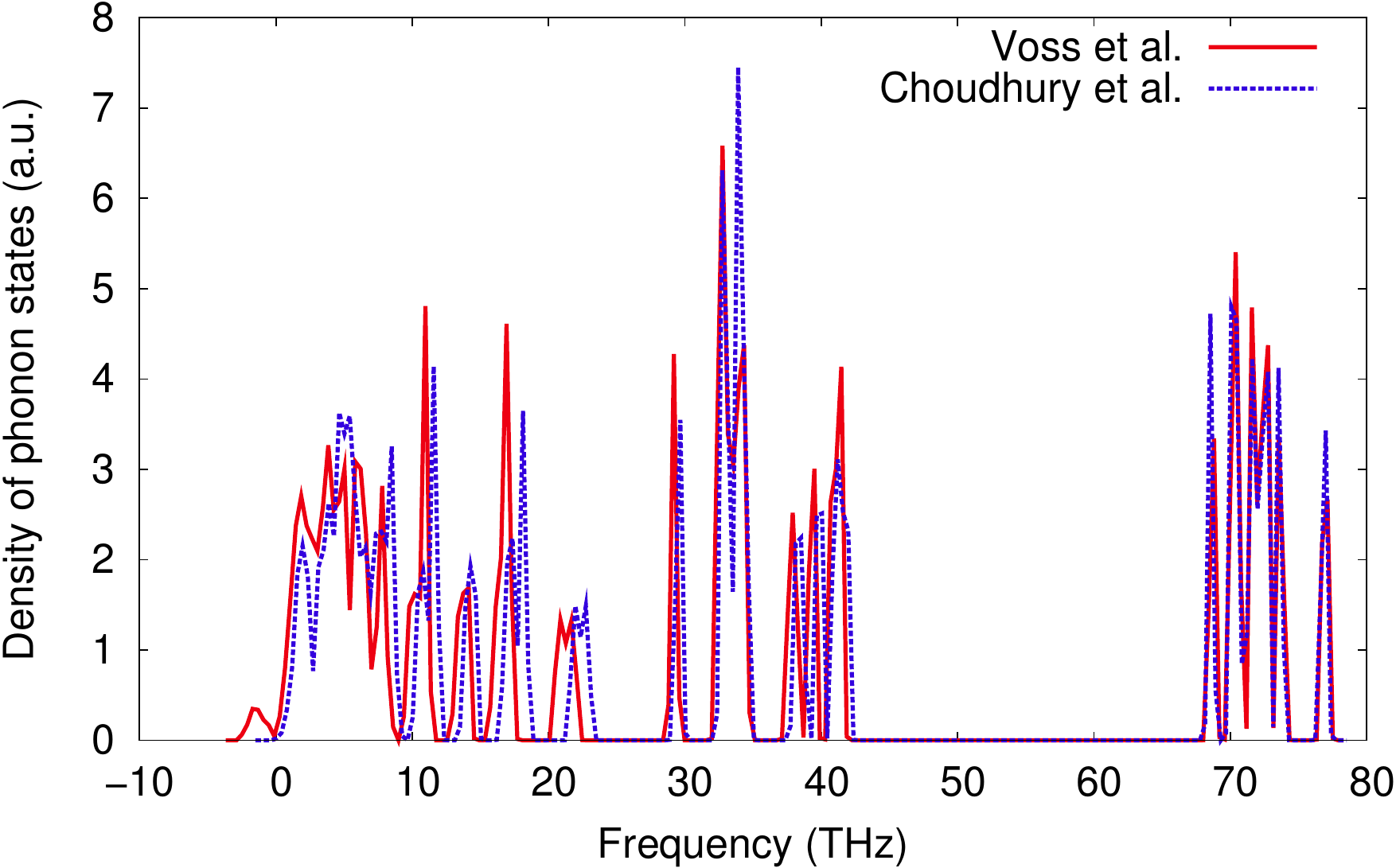}
  \caption{(Color online) Density of phonon states of the $Pmc2_1$ structures for Mg(BH$_4$)$_2$ taken from Voss {\it et al.}, J. Phys. Condens. Matter {\bf 21}, 012203 (2009) --- solid curve, and from Choudhury {\it et al.}, Phys. Rev. B {\bf 77}, 134302 (2008) --- dashed curve.}\label{fig:phonondos_sg26}
  \end{center}
\end{figure}
\newpage

\begin{figure}[ht]
  \begin{center}
    \includegraphics[width=10cm]{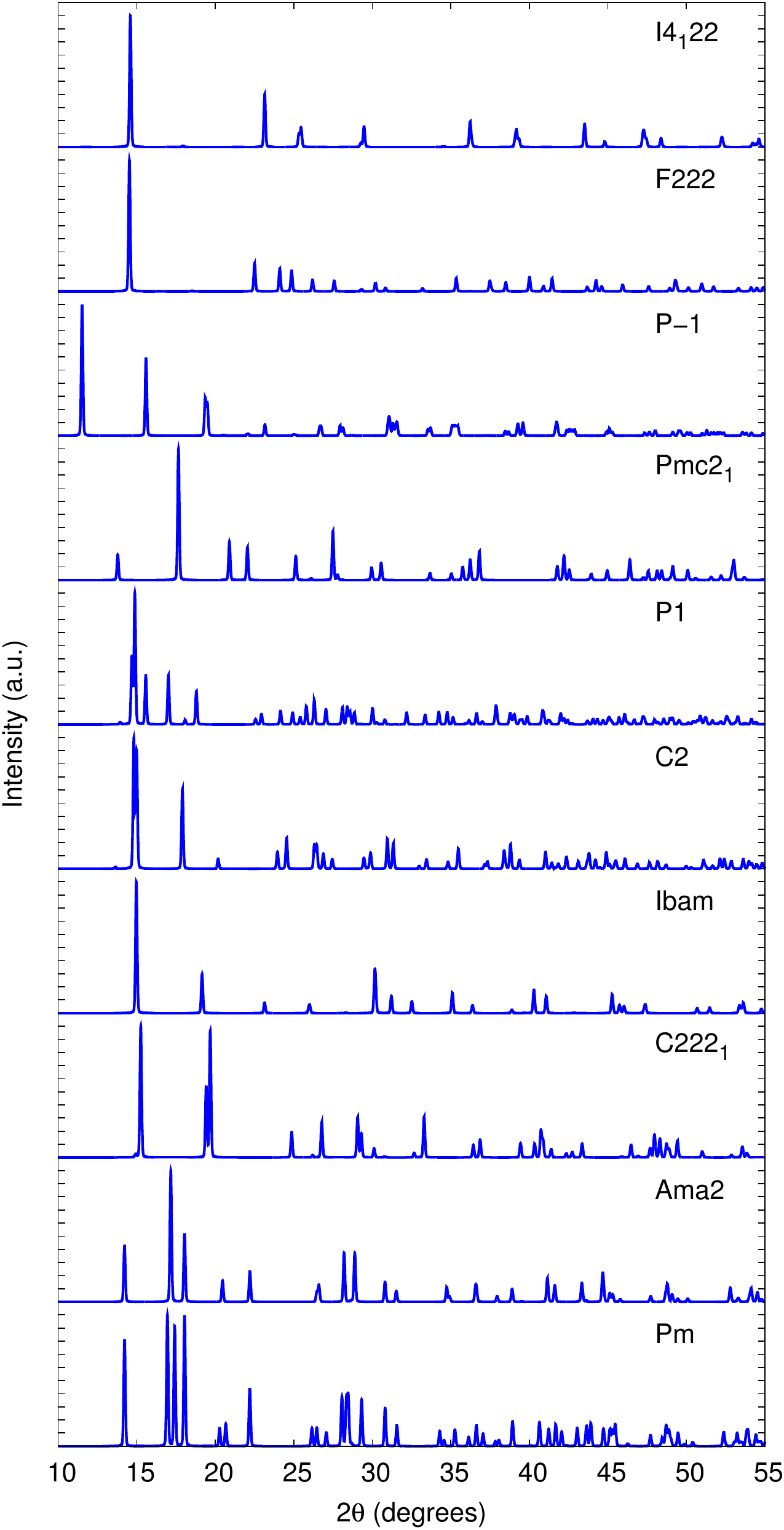}
  \caption{(Color online) Calculated XRD patterns of the dynamically stable structures for Zn(BH$_4$)$_2$.}\label{fig:xrd}
  \end{center}
\end{figure}